\documentclass[preprint,12pt]{elsarticle}

\usepackage{amssymb}
\usepackage{amsmath}

\begin{document}

\begin{frontmatter}



\title{Degree of diffraction for monochromatic light beams}

\author{Jie Wu}
\ead{jiewu@shu.edu.cn}

\author{Shuang-Yan Yang}
\ead{shuangyan@shu.edu.cn}

\author{Chun-Fang Li\corref{CA}}
\ead{cfli@shu.edu.cn}

\cortext[CA]{Corresponding author}

\address{Department of Physics, Shanghai University, 99 Shangda Road, Shanghai 200444, China}

\begin{abstract}
A parameter, called the degree of diffraction, is defined to describe the diffractive spreading of a monochromatic light beam. The same as the degree of paraxiality that was introduced by Gawhary and Severini in Opt. Lett. 33, 1360 (2008), the degree of diffraction depends only on beam's angular spectrum. With this definition, it is possible to quantitatively compare the diffractive spreading of different light beams.

\vspace{10pt}


\end{abstract}


\end{frontmatter}


\section{Introduction}

One of propagation properties of light beams is their diffraction behavior. It is well known that for a fundamental Gaussian beam of field distribution
\begin{equation}\label{GB}
E_G (\rho, \phi) \propto \exp \bigl(-\frac{\rho^2}{w^2_0}  \bigr)
\end{equation}
at waist plane $z=0$, a propagation distance,
$L_R=\pi w^2_0/\lambda$,
called the Rayleigh range \cite{Siegman}, is used to characterize its diffraction behavior, where $\lambda$ is the wavelength.
For a truncated or pseudo diffraction-free beam \cite{Durnin, Durnin-ME, Gori, Turunen, Rosen, Rosen-SY, Bouchal, Turunen-F}, a propagation distance called the diffraction-free range is advanced. As an example, for a Bessel-Gaussian (BG) beam \cite{Gori, Turunen-F} of field distribution
\begin{equation}\label{BGB}
E_{BG} (\rho, \phi) \propto J_0 (\beta \rho) \exp \bigl(-\frac{\rho^2}{w^2_0} \bigr)
\end{equation}
at $z=0$, the diffraction-free range is \cite{Gori}
$L_G=2 \pi w_0/(\beta \lambda)$,
where condition
\begin{equation}\label{condition-Beta}
    \frac{\beta w_0}{2} >1
\end{equation}
should be satisfied \cite{Gori} for the BG beam to behave like a portion of the diffraction-free beam.
The so-called Rayleigh range \cite{Hodgson-W} for a truncated Bessel beam is in fact the diffraction-free range \cite{Durnin, Durnin-ME}.

Although the diffraction-free range of the BG beam is much shorter \cite{Porras} than the Rayleigh range of its component Gaussian beams (according to Gori {\it et. al.} \cite{Gori}, a BG beam is produced by the superposition of Gaussian beams whose axes are uniformly distributed on a cone),
$L_G=\frac{L_R}{\beta w_0/2} < L_R$,
it is hard to quantitatively compare the diffractive spreading of the Gaussian beam (\ref{GB}) with that of the BG beam (\ref{BGB}). This is because physically the Rayleigh range of the Gaussian beam is defined differently from the diffraction-free range of the pseudo diffraction-free beam. The Rayleigh range is defined \cite{Siegman} as the distance along the propagation direction of the Gaussian beam from its waist to the point where the area of the cross section is doubled. On the other hand, the diffraction-free range is defined \cite{Durnin, Durnin-ME} as the propagation distance over which the profile of the pseudo diffraction-free beam remains invariant. To the best of our knowledge, there does not exist a universal parameter to describe the diffractive spreading of light beams. Even the commonly used $M^2$ factor, the beam-quality factor, does not meet the need. As is well known, Gaussian beams with different waist radii $w_0$ have different Rayleigh ranges. But they all have the same $M^2$ factor, the unity.
The purpose of this paper is to introduce such a parameter.

\section{Degree of diffraction is different from degree of paraxiality}

To this end, let us first make use of diffraction-free beams to show that the diffraction behavior of a light field is conceptually independent of its another propagation property, the paraxiality \cite{Gawhary-OL, Gawhary-OC, Wang}. It is known that the wavevector of all the plane wave that composes a diffraction-free beam lies on a cone. In the scalar case \cite{Durnin}, the angular spectrum can be written as \cite{Turunen-F}
\begin{equation*}
    f(\mathbf{w})= A(\varphi) \delta(k_\rho -\beta)
\end{equation*}
in circular cylindrical coordinates, where
$\mathbf{w}={\mathbf k}/k$
is the unit wavevector,
$\delta$ is the Dirac delta function,
$\beta=k \sin \vartheta_0$,
and $\vartheta_0$ denotes the apex half-angle of the wavevector cone. In the vectorial case \cite{Bouchal-O, Wang-11}, the angular spectrum can be factorized into a polarization vector $\mathbf{e} (\mathbf{w})$ and the above scalar angular spectrum,
\[
\mathbf{f} (\mathbf{w})= \mathbf{e}(\mathbf w) A(\varphi) \delta(k_\rho -\beta),
\]
where the polarization vector is a unit vector,
$|\mathbf{e}|=1$,
and is constrained by the transversality condition
$\mathbf{e} \cdot \mathbf{w}=0$.
In both cases the paraxiality of the diffraction-free beam is completely determined \cite{Gawhary-OL, Wang} by the apex half-angle $\vartheta_0$ of the wavevector cone. The smaller the angle $\vartheta_0$ is, the larger the degree of paraxiality is.
Nevertheless, as the name suggests, all the diffraction-free beams with different $\vartheta_0$ should have the same diffraction behavior. They are all free of diffraction. This means that the diffraction behavior of a beam is indeed conceptually independent of its paraxiality.
Therefore it is essential to introduce a parameter that is different from the degree of paraxiality to characterize the diffraction behavior. Such a parameter will be referred to as the degree of diffraction (DOD).

To introduce the DOD, it is instructive to analyze the mechanism for the diffractive spreading of a light beam. For this purpose, let us look at the difference between a diffractive beam and a diffraction-free beam in the angular spectrum.
As mentioned before, the wavevector in a diffraction-free beam is only distributed on a cone, which means that a diffraction-free beam is an eigen state of the longitudinal component of the momentum \cite{Turunen-F}. So the standard deviation of $k_z$ in this case is equal to zero,
\begin{equation}\label{VD}
    \Delta k_z \equiv [\langle k_z^2 \rangle- \langle k_z \rangle^2]^{1/2}=0,
\end{equation}
where
\begin{equation*}
\langle Q \rangle=\frac{\int |\mathbf{f}|^2 Q d \Omega}{\int |\mathbf{f}|^2 d \Omega}
                 =\frac{\int |f|^2 Q d \Omega}{\int |f|^2 d \Omega}
\end{equation*}
is the expectation value \cite{Gawhary-OL, Gawhary-OC} of quantity $Q$, $\mathbf f$ is the vector angular spectrum, and
$d \Omega= \sin \vartheta d \vartheta d \varphi$
is the solid-angle element in wavevector space.
For a diffractive beam, we take the BG beam as an example, which can be regarded as composing of Gaussian beams whose axes are uniformly distributed on a cone \cite{Gori}.
Because the angular spectrum of a Gaussian beam also takes a Gaussian form, it follows that the wavevector in a BG beam is distributed around the axis cone. Consequently, the standard deviation of $k_z$ does not vanish.
This shows that whether a light beam is diffraction-free or not depends on whether the standard deviation of its $k_z$ vanishes or not.
Considering that any coherent light field can be expanded in terms of a particular complete set of diffraction-free beams \cite{Bouchal-O, Wang-11}, the diffractive spreading of a light beam propagating in free space lies in the interference between its diffraction-free components that have different wavevector cones.

\section{Definition of DOD}

The same as the degree of paraxiality \cite{Gawhary-OL, Gawhary-OC, Wang}, a well defined DOD should also range from 0 to 1. To see how to define such a DOD on the basis of the above analysis, we consider two extreme light fields. One is the least diffractive fields, the diffraction-free beams.
Just as its name implies, the DOD for such beams should be equal to 0. In view of this, the expected DOD should be proportional to $\Delta k_z$ in accordance with Eq. (\ref{VD}).
The other extreme light field is described by angular spectrum of vector spherical harmonics,
$\mathbf{f}(\mathbf w)= \mathbf{e}(\mathbf w) Y_{\lambda \mu} (\mathbf w)$,
where
\[
    Y_{\lambda \mu} (\mathbf{w})
   =\left\{ \frac{2\lambda +1}{4 \pi} \frac{(\lambda-\mu)!}{(\lambda+\mu)!} \right\}^{1/2}
    P_{\lambda}^{\mu} (\cos \vartheta) e^{i \mu \varphi}
\]
is the spherical harmonics satisfying normalization condition
$\int |Y_{\lambda \mu}|^2 d \Omega =1$.
Due to the spherically symmetric distribution of the wavevector, such a field is the most ``diffractive''. The DOD for this field should be equal to 1. Because the expectation value of $k_z$ in this field vanishes,
$\langle k_z \rangle=0$,
as can be easily checked, we have
$\Delta k_z=\langle k_z^2 \rangle^{1/2}$.
Combining these two considerations together, it seems reasonable to define the DOD for any coherent light field as
\begin{equation}\label{D'}
    D'=\frac{\Delta k_z}{\langle k_z^2 \rangle^{1/2}}
      =\Big( 1-\frac{\langle k_z \rangle^2}{\langle k_z^2 \rangle} \Big)^{1/2}.
\end{equation}
So defined DOD has the following properties:
\begin{enumerate}
  \item The same as the degree of paraxiality \cite{Gawhary-OL, Gawhary-OC, Wang}, it does not necessarily require the knowledge about the intensity distribution of the electric field in spatial space. Besides, it does not depend on the vector nature of the angular spectrum and therefore applies to scalar as well as vector fields.
  \item As is required, it ranges from 0 to 1. For the least diffractive fields, the diffraction-free beams, it is equal to 0; for the most diffractive fields such as the spherical waves, it is equal to 1.
  \item It describes the diffractive spreading of a beam in such a way that the more diffractive a beam is, the larger its value is.
\end{enumerate}

It is noted that $D'$ in Eq. (\ref{D'}) is proportional to the standard deviation of $\cos \vartheta$,
$\sigma_P= (\langle \cos^2 \vartheta \rangle-\langle \cos \vartheta \rangle^2)^{1/2}$,
that was discussed in Ref. \cite{Gawhary-OC}. Indeed, with the help of $k_z=k \cos \vartheta$, Eq. (\ref{D'}) can be written as
\[
    D'=\frac{\sigma_P}{\langle \cos^2 \vartheta \rangle^{1/2}}.
\]
Noticing that the degree of paraxiality introduced in Ref. \cite{Gawhary-OC} is
$\langle \cos \vartheta \rangle$,
which is different from the denominator
$\langle \cos^2 \vartheta \rangle^{1/2}$,
the ratio
$\sigma_P/\langle \cos \vartheta \rangle$
discussed there is something like, but different from, $D'$.
Let us appreciate the above mentioned third property of $D'$ in the following by applying it to a uniformly distributed angular spectrum with respect to the polar angle.

\begin{figure}[htb]
\centerline{\includegraphics[width=5.5cm]{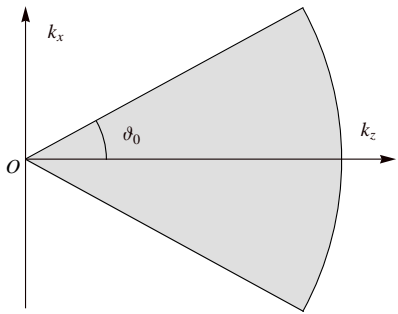}}
\caption{Schematic illustration of the cone within which the wavevector is uniformly distributed.} \label{AS}
\end{figure}

\begin{figure}[htb]
\centerline{\includegraphics[width=5.5cm]{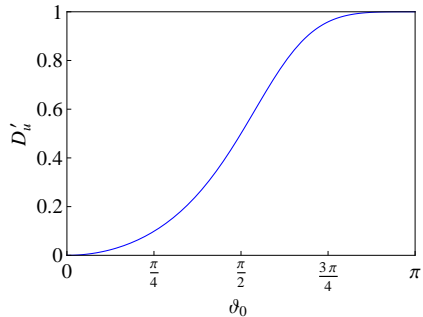}}
\caption{Dependence of $D'_u$ on beam parameter $\vartheta_0$.} \label{Du}
\end{figure}

Since $D'$ in Eq. (\ref{D'}) does not depend on the vector nature of the angular spectrum, the angular spectrum that we consider assumes the form,
\begin{equation}\label{AS-UD}
    f_u (\mathbf w)
   =\begin{cases}[2 \pi (1-\cos \vartheta_0)]^{-1/2}, & 0 \leq \vartheta<\vartheta_0,    \\
                              0,                      & \vartheta_0 < \vartheta \leq \pi,
    \end{cases}
\end{equation}
which satisfies normalization condition
$\int |f_u|^2 d \Omega=1$.
It describes a beam the wavevector of which is uniformly distributed within a cone of apex half-angle $\vartheta_0$, as is illustrated graphically in Fig. \ref{AS}.
When $\vartheta_0 =0$, it describes a plane wave propagating in the $z$ direction. The DOD in this case should be equal to 0. With the increase of $\vartheta_0$, the DOD should increase monotonously.
When $\vartheta_0$ approaches $\pi$, we will arrive at a beam the wavevector of which is uniformly distributed in the whole solid angle of $4 \pi$. This is one of the most diffractive beams and should have a DOD of unity.
Such a feature is well reflected by the DOD (\ref{D'}). In fact, straightforward calculations with Eq. (\ref{D'}) give
\begin{equation}\label{D-u}
    D'_u=\frac{1}{2}
         \Big( 1-\frac{3 \cos \vartheta_0}{1+\cos \vartheta_0 +\cos^2 \vartheta_0} \Big)^{1/2}.
\end{equation}
The dependence of $D'_u$ on $\vartheta_0$ is shown in Fig. \ref{Du}. It properly reveals the above expectation.

As can be seen from Eq. (\ref{D-u}), when $\vartheta_0=\pi/2$, that is to say, when no component plane waves propagate in the negative $z$ direction, one has $D'_u=1/2$. If we further require that the DOD be applied only to beam-like light fields, that is to say, only to those light fields no component plane waves of which propagate in the backward direction, and that the DOD of the most diffractive beam in this case be still equal to unity, definition (\ref{D'}) should be replaced with
\begin{equation}\label{DOD}
    D=2 D' =2 \Big( 1-\frac{\langle k_z \rangle^2}{\langle k_z^2 \rangle} \Big)^{1/2}.
\end{equation}
Equation (\ref{DOD}) is the primary result of this paper.
It should be emphasized that so defined DOD is a beam parameter, having nothing to do with the propagation distance. Let us make use of it to discuss two examples below.

\section{Applications}

\subsection{Gaussian-like beams}

\begin{figure}[htb]
\centerline{\includegraphics[width=5.5cm]{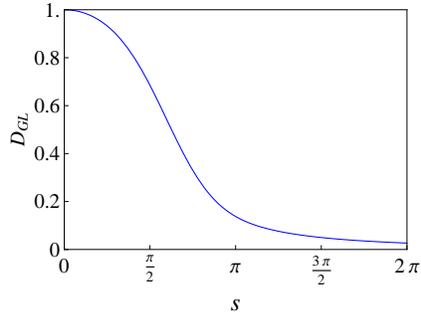}}
\caption{Dependence of $D_{GL}$ on the beam parameter $s$.} \label{Dgl}
\end{figure}

In the first example, we consider a Gaussian-like beam \cite{Wang} that has the following angular spectrum,
\begin{equation}\label{AS-GL}
    f_{GL}=A_0 \exp\Big( -\frac{k^2 w_0^2}{4} \sin^2 \vartheta \Bigr),
           \quad 0 \leq \vartheta \leq \frac{\pi}{2}.
\end{equation}
It is normalized in accordance with
\begin{equation}\label{normalization}
    \int |f_{GL}|^2 d\Omega=2 \pi A_0^2 \frac{F(s)}{s},
\end{equation}
where
$s=k w_0/\sqrt{2}$ and
\[
    F(x)=\exp(-x^2) \int_0^x \exp(t^2) dt
\]
is the Dawson function. With the help of Eq. (\ref{normalization}), one easily obtains
\[
    \langle k_z \rangle=\frac{1-\exp(-s^2)}{2sF(s)} k
\]
and
\[
    \langle k_z^2 \rangle=\frac{s-F(s)}{2 s^2 F(s)} k^2.
\]
Substituting them into Eq. (\ref{DOD}) gives
\begin{equation}\label{D-GL}
    D_{GL}=2 \Big\{ 1-\frac{[1-\exp(-s^2)]^2}{2F(s)[s-F(s)]} \Big\}^{1/2}.
\end{equation}
When $s \rightarrow 0$, angular spectrum (\ref{AS-GL}) reduces to the uniformly distributed angular spectrum (\ref{AS-UD}) with
$\vartheta_0 =\pi/2$.
In this case, we have
$D_{GL} \rightarrow 1$, which is just what we expect.
In obtaining this result, we have made use of the following Taylor series of the Dawson function,
\[
    F(x)=\sum_{n=0}^\infty \frac{(-1)^n 2^n}{(2n+1)!!} x^{2n+1}=x-\frac{2}{3} x^3+\cdots
\]
To the opposite, when $s \rightarrow \infty$, we find
$D_{GL} \rightarrow 0$, where the asymptotic behavior, $F(x)\sim 1/(2x)$, of the Dawson function at large $|x|$ is used. This is also what we expect, because the beam in this case approaches a plane wave, a diffraction-free beam. With the increase of $s$, the DOD decreases monotonously from 1 to 0. The dependence of $D_{GL}$ on the beam parameter $s$ is schematically shown in Fig. \ref{Dgl}.

It is worth noting \cite{Chen} that when $s$ is large enough
(taking into account the difference between our expression (\ref{AS-GL}) for the angular spectrum and Eq. (8) in \cite{Chen}, the beam parameter $s$ here is exactly the reciprocal of the beam parameter $f$ in \cite{Chen}), for instance, when
\begin{equation}\label{large-s}
    s>2 \pi,
\end{equation}
angular spectrum (\ref{AS-GL}) describes a beam that has a negligible longitudinal component (only about $1\%$ of the transverse component in intensity) and can be approximated by the scalar Gaussian beam (\ref{GB}).

\subsection{BG beams}

In the second example, we compare quantitatively the diffractive spreading of the BG beam with that of its component Gaussian beams. The angular spectrum of the BG beam (\ref{BGB}) is given by \cite{Turunen-F}
\[
    f_{BG}=I_0 \Bigl( \frac{\beta w_0^2}{2} k_\rho \Bigr)
           \exp \Bigl[-\frac{w_0^2}{4} (\beta^2 +k_\rho^2) \Bigr]
\]
in circular cylindrical coordinates, where $I_0$ is the modified Bessel function of the first kind and zeroth order. Because the evanescent components that correspond to $k_\rho >k$ do not contribute to the diffraction, we convert this angular spectrum into
\begin{equation}\label{AS-BG}
    f_{BG}=I_0 \Bigl( \frac{k\beta w_0^2}{2} \sin\vartheta \Bigr)
           \exp \Bigl[-\frac{w_0^2}{4} (\beta^2 +k^2 \sin^2 \vartheta) \Bigr]
\end{equation}
in spherical coordinates, where
$0 \leq \vartheta \leq \pi/2$.

Although it is hard to obtain an analytical expression for the DOD, $D_{BG}$, of the BG beam (\ref{AS-BG}), we can still calculate numerically the dependence of $D_{BG}$ on the beam parameter $s$. Furthermore, when $\beta=0$, Eq. (\ref{AS-BG}) reduces to Eq. (\ref{AS-GL}). In this case, we arrive at the Gaussian beam at large $s$. This shows that definition (\ref{DOD}) for the DOD allows us to compare quantitatively the diffractive spreading of the BG beam with that of its component Gaussian beams.
\begin{figure}[htb]
\centerline{\includegraphics[width=5.5cm]{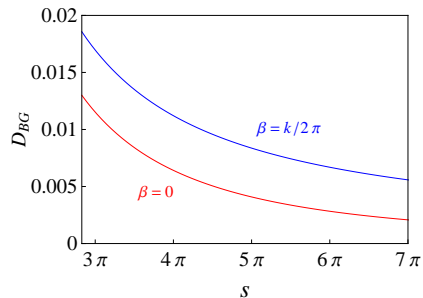}}
\caption{Comparison of the DOD of the BG beam ($\beta=k/2\pi$) with that of its component Gaussian beams ($\beta=0$).} \label{Dbg}
\end{figure}
In Fig. \ref{Dbg} is shown such a comparison, where the blue curve is for the DOD of the BG beam, and the red curve is for the DOD of the component Gaussian beams. The parameters for the BG beam are chosen as follows. Letting the radius $\rho_0$ of the central spot of the Bessel factor be about the wavelength, $\rho_0=\lambda$, we have \cite{Durnin-ME} $\beta= 1/\rho_0 =k/2\pi$.
Moreover, in order that condition (\ref{condition-Beta}) be satisfied, we choose $s>2 \sqrt{2} \pi$, which also meets the requirement (\ref{large-s}). Clearly, the BG beam is more diffractive than its component Gaussian beams.
It should be pointed out that the comparison made here is not to be confused with the comparison that was made in the literature \cite{Durnin, Durnin-ME, Hodgson-W} between a truncated Bessel beam and a Gaussian beam. After all, a truncated Bessel beam cannot be viewed as composing of Gaussian beams.

\section{Conclusions}

In conclusion, we introduced a parameter called the DOD to describe the diffractive spreading of light beams. The same as the degree of paraxiality \cite{Gawhary-OL, Gawhary-OC, Wang}, the DOD depends only on the angular spectrum of light beams. It ranges from 0 to 1. It describes the diffractive spreading of a light beam in such a way that the more diffractive a light beam is, the larger its value is. Specifically, it is equal to zero for the diffraction-free beams. Besides, we pointed out the relation of the DOD with the standard deviation \cite{Gawhary-OC} of the cosine of the angle that the wavevector makes with the beam axis.
The parameter of DOD should be useful in areas in which the assessment of beam quality, especially its diffraction spreading, is of importance, such as in laser processing \cite{Bhuyan} and in free space optical communication \cite{Jia}.

\section*{Acknowledgements}

This work was supported in part by the National Natural Science Foundation of China (61108010).



\end{document}